# The electronic, thermodynamic, thermoelectric and optical properties of Ca(InP)$_2$ compound: DFT study

S. Dahri [1], A. Jabar [1,2], L. Bahmad [3,*], L. B. Drissi [1], and R. Ahl Laamara [1]

[1] LPHE-MS, Science Faculty, Mohammed V University in Rabat, Morocco

[2] LPMAT, Faculty of Sciences Aïn Chock, Hassan II University of Casablanca, B.P. 5366 Casablanca, Morocco

[3] Laboratory of Condensed Matter and Interdisciplinary Sciences (LaMCScI), Faculty of Sciences, Mohammed V University, Av. Ibn Batouta, B. P. 1014 Rabat, Morocco

*Corresponding author: l.bahmad@um5r.ac.ma (L.B.)

**Abstract**

In this study, we investigate the electronic, optical, thermoelectric, and thermodynamic properties of Ca(InP)$_2$ through comprehensive theoretical calculations Ca(InP)$_2$ is a compound with promising applications in materials science and electronics. Using the density functional theory (DFT) with Generalized Gradient Approximation (GGA) and modified Becke–Johnson approximation (mBJ), we determine the band structure, density of states, and optical properties of Ca(InP)$_2$. The obtained results reveal that the Ca(InP)$_2$ compound exhibits a direct band gap of 0 eV and 0,645 eV for PBE-GGA and GGA+mBJ, respectively. This direct band gap is found at the $\Gamma$ point of the Brillouin zone, making it well-suited for optoelectronic applications. Furthermore, we analyze the thermoelectric properties such as the Seebeck coefficient, the lattice thermal conductivity, and optical properties like dielectric function, absorption coefficient, conductivity, and extinction coefficient. Thermodynamic properties, including heat capacity and Debye temperature, are also calculated, providing a deeper understanding of the compound's thermal behavior. The findings of this study highlight the fundamental characteristics of Ca(InP)$_2$ and offer valuable information for its potential use in electronic and optoelectronic devices. A comprehensive understanding of the electronic, optical, and thermodynamic properties of the Ca(InP)$_2$ compound can serve as a guide for future experimental research and aid in the design of novel materials for a wide range of technological applications.

**Keywords:** Ca(InP)$_2$; GGA+mBJ; Electronic properties; Thermoelectric properties; Optical properties; Thermodynamic properties.

## 1. Introduction

Ca(InP)$_2$ is a significant example of two-dimensional materials that possess distinct electronic and optical properties, making them promising for a variety of technological applications [1,2]. Its regular hexagonal crystal structure, usually obtained by adjusting growth conditions and manufacturing parameters, offers considerable potential for applications ranging from advanced electronic devices to integrated sensors and optoelectronic devices [3-6]. Two-dimensional materials, such as Ca(InP)$_2$, often feature Zintl phases, characterized by complex structures resulting from the combination of ionic and covalent bonds [7-9]. These Zintl phases add an extra dimension to their electronic behavior, enabling versatile use in electronics and semiconductor technology [10]. Ca(InP)$_2$ crystallizes in the hexagonal P6$_3$/mmc space group (Number 194), the lattice parameters of Ca(InP)$_2$ a=b=4.05 Å, c=17.49 Å and α = β =90°, γ =120° [11]. The structure consists of two Ca(InP)$_2$ sheets aligned in the (0, 0, 1) direction. The electron configuration for calcium (Ca): [Ar] 4s$^2$ is alkaline earth metal [12], indium (In): [Kr] 4d$^{10}$ 5s$^2$ 5p$^1$ is poor metal [13-15], and phosphorus (P): [Ne] 3s$^2$ 3p$^3$ it is non-metal [16]. The typical difference between materials represents different electronegativity values between atoms with high electronegativity for calcium 1.00 and phosphorus 2.19, indicating a transfer of electrons from calcium to phosphorus [17]. The bond between calcium and phosphorus in the compound Ca(InP)$_2$ is an ionic bond. In an ionic bond, electrons are transferred completely from one atom to another, forming charged ions. In this case, calcium (Ca) tends to lose two valence electrons to form a Ca(InP)$_2$ ion, while phosphorus (P) tends to gain three valence electrons to become a P$^{3-}$ ion [18]. Indium (In) and phosphorus have a relatively small difference in electronegativity, indicating a tendency to share electrons rather than transfer them completely. Indium has an electronegativity of around 1.78. this moderate difference in electronegativity suggests that the band between indium and phosphorus is predominantly covalent [19,20], with some polarity in the bond. The valence electrons of the two atoms are shared, forming a binding electron pair that holds the atoms together [21]. The bond linking calcium and indium in the compound Ca(InP)2 is a metallic bond [22], with the valence electrons of the metal atoms liberated to form a common "electron cloud" shared by the atoms.

This work is structured as follows: section 2 describes the ab initio method. Then, in section 3, we delve into the analysis of electronic properties and examine thermodynamic properties. We then turn to thermoelectric properties, including formulas for calculating physical quantities using Wien2k. In addition, we explore the optical properties and present the numerical results and corresponding discussion. Finally, section 4 outlines the conclusions drawn from this study.

## 2. Computation details

In principle, the study of material properties involves describing the behavior of a set of interacting electrons and nuclei and trying to understand how the arrangement of atoms and the way they interact give this set its particular properties. For a fundamental understanding of these properties, several computational methods have been developed. One of the most rigorous and sophisticated is the ab initio method [23]. Based on fundamental quantum theory, ab initio methods have become an indispensable tool for predicting and studying the structural, electronic, mechanical and optical properties of materials. Among ab initio methods, Density Functional Theory (DFT) [24,25] is one of the most widely used for quantum calculations of the electronic structure of matter (atoms, molecules, solids), due to the drastic simplification it brings to the solution of the Schrödinger equation, which is based on the consideration that the total energy of a system is a function of its electronic density. To calculate the equilibrium structural parameters and electronic band structure of $Ca(InP)_2$, we used the Wien2k [26]. In this work, we are interested in the study of the properties of $Ca(InP)_2$ using the generalized gradient approximation (PBE-GGA) [27] method and the modified Becke-Johnson approximation (GGA+mBJ). The specific number of k points used in the irreducible integral of the Brillouin zone is 10x10x10, and the density value of the cut-off plane is $RK_{max} = 9$. Firstly, to find out the energy value at which the $Ca(InP)_2$ structure is stable and discover the new parameters after optimization with constant a: b: c ratio. The specific RMT values for calcium (Ca), indium (In) and phosphorus (P) play an essential role in the precise determination of the $Ca(InP)_2$, as well as in the overall performance of electronic calculations within Wien2k. The RMT values Ca, In and P are 2.5, 2.49 and 2.04 respectively, which seems to indicate a significant difference in the radii of these atoms, reflecting their distinct size characteristics in the crystal structure. By adjusting these RMT values, calculations in Wien2k take into account the specific geometry of the crystal and the size of the constituent atoms to obtain an accurate representation of the electronic wave functions within each Muffin-Tin sphere. The difference in RMT values between Ca, In and P suggests distinct electronic interactions in the corresponding Muffin-Tin spheres, which may influence the band gap, surface states and other important electronic properties of the Ca(InP)$_2$ material. And is to calculate thermodynamic properties to understand how they react under pressure range varied from P= 0 GPa to P= 12 GPa or temperature between T= 0K and T= 1800K conditions we have employed a Gibbs calculation program [28].

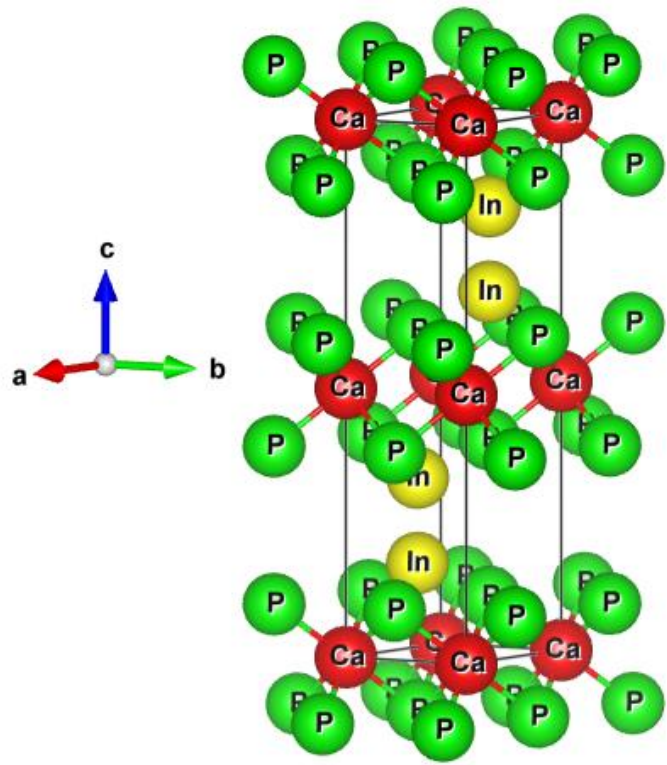

*Fig. 1: Unit-cell crystalline structure of the layered Zintl phase Ca(InP)$_2$.*

### 3. Results and discussions

### 3.1. Total energy and lattice parameters

we performed a self-consistent calculation of the total energy by varying the lattice parameter. The values of these parameters were chosen around the experimental lattice parameter. The equilibrium lattice parameter was calculated by fitting the total energy curve obtained using the Birch-Murnaghan equation of state [29], which is given by:

$$E(V) = E_0 + \left[\frac{B_0 V}{B'(B'-1)}\right]\left[B'\left(1 - \frac{V_0}{V}\right) + \left(\frac{V_0}{V}\right)^{B'} - 1\right] \qquad (1)$$

$$E = E_0 + \frac{9}{16}\left(\frac{B}{14703.6}\right)V_0[(\eta^2 - 1)^3 B' + (\eta^2 - 1)^2(6 - 4\eta^2)] \qquad (2)$$

And

$$\eta = \left(\frac{V_0}{V}\right)^{1/3}$$

Where $V_0$ is the volume of the unit mesh in the fundamental state. The modulus of compressibility (B) and its derivative B' are, in turn, related to the volume of the unit mesh where (B) is a physical quantity that measures the response of a material to uniform compression. It quantifies the change in volume of a material under the influence of an applied pressure. More precisely, the modulus of inertia measures a material's rigidity or resistance to compression. We optimize using c/a ratio optimization for hexagonal and lattices, the optimized parameter is also shown in Table 1. A unit cell of Ca(InP)$_2$ has positions of Ca (0, 0, 1/2), In (1/3, 2/3, 0.670) and P (2/3, 1/3, 0.602) respectively. Figure 2 shows the variation of energy as a function of unit cell volume for the hexagonal structure of the compound Ca(InP)$_2$. The minimum energy is -52525.256 Ry and the minimum volume is 970.601 Å$^3$, as determined using the GGA-PBE potential approach.

**Table 1:** Values of $a_0$ (in Å), $V_0$ (in Å$^3$), B (in GPa), B', and $E_0$ (in Ry) by using the PBE-GGA approach.

| $a_0$(Å) | $c_0$(Å) | $V_0$(Å$^3$) | B(GPa) | B' | $E_0$(Ry) |
|---|---|---|---|---|---|
| 4.07 | 18.65 | 253.894 | 70.910 | 16.063 | -52525.256 |

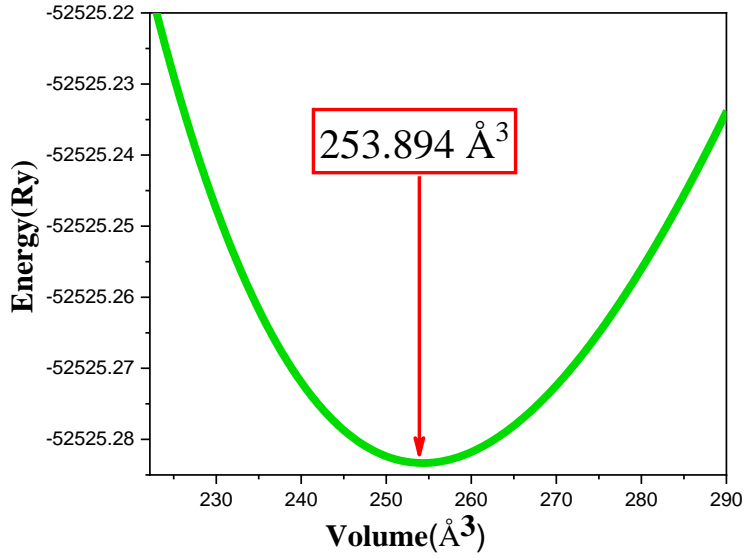

*Fig.2:* *The energy vs the volume of the Ca(InP)₂.*

### 3.2. Electronic properties

The calculation of electronic properties of the explored materials, including the Band structure, the total density of states (TDOS) with a partial density of states (PDOS)curves, and the calculation of the electronic band structure provides information on the electronic wave functions at wave number and a certain energy. First, the electronic structures for $Ca(InP)_2$ were calculated by PBE–GGA and GGA+mBJ potentials. shown in Figure 3, The left part (a) displays the electronic band structure using the Perdew–Burke–Ernzerhof PBE-GGA potential, while the right part (b) illustrates the electronic band structure using the GGA+mBJ. Represents the band structure curve of the $Ca(InP)_2$ nanosheet in symmetrical in the first Brillouin zone, which indicates that the direct path deviation at point A by PBE-GGA and $\Gamma$ by GGA+mBJ, for the PBE-GGA approximation it is conductive because the energy gap between the maximum of the valence band and the minimum of the conduction band is $E_g=0$ eV. However, with the GGA+mBJ approximation, we observe a gap energy value of $E_g=0.59$ eV, this indicates that it is a p-type semiconductor [30] with a direct band gap. According to the experimental band gap value of these materials [7], we have found that $Ca(InP)_2$ is a semiconductor. Therefore, the PBE-GGA approximation gives an imprecise value for the energy gap. To address this, we apply compression to $Ca(InP)_2$ along the z-axis and then observe the variation of the energy gap as a function of the lattice parameter $c_0$. We noticed an increase in the energy gap value with compression along the z-axis, that is, a decrease in the lattice parameter $c_0$, up to the value of

0.222 eV. This new value, the band gap, is with which we will complete the study of the remaining properties of this substance.

The calculated total density of state (TDOS) as a function of the energy projected between -2 eV and 2 eV for Ca(InP)$_2$ is shown in Fig. 5 and is calculated by the PBE-GGA and GGA+mBJ approximations. As we see, this is symmetric concerning the axis of energy, it shows that material Ca(InP)$_2$ is not magnetic. For both approximations, Ca(InP)$_2$ is a semiconductor with a difference in Gap energy value. Moreover, the band Gap of TDOS is located around 1 eV for PBE-GGA and 1.5 eV for GGA+mBJ. For the PDOS (partial density of state), the phosphorus atom exerts a dominant influence on the density of state, followed by indium and calcium were shown in Fig. 6 (a), the decomposition of TDOS between - 2 eV and 2eV for the three atoms, firstly for the calcium atom we note that the valence band of Ca is equally composed of two energy ranges, the first between -2eV and -1.5 eV, where the d-state of Ca has a major contribution, and the second between 0.75 eV and 2 eV also shows a substantial contribution from the d-state (see Fig.6(b)) with a small contribution from p-state. For the P atom well, the valence band is also composed of two energy ranges the first between -2 and 0 eV and the second between 0.5 eV and 2 eV the band is formed mainly by the p-state with a low contribution from the s-state. And for the last atom of In is shown in Fig. 6 (c) the valence band is composed of three energy ranges between -2 and 0 eV and 1.3 eV and 2 eV the p-state has a major contribution and for the energy range between 0.5 eV and 1.3 eV the band formed by s-state with a low contribution of p-state. We note that the valences band of Ca(InP)$_2$ Alloy is also composed of two energy ranges

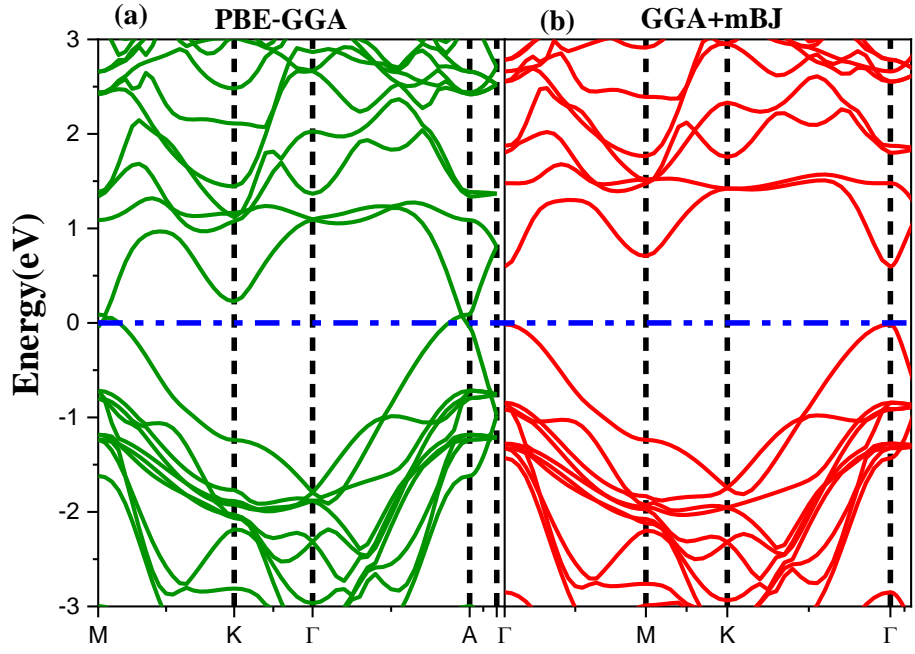

*Fig.3: (a) Band structures of the Ca(InP)₂ using PBE -GGA and (b) Band structures of the Ca(InP)₂ using GGA+mBJ.*

**Table. 2:** the variation of the Gap energy as a function of the lattice parameter $c_0$.

| $c_0$ | 15.65 | 16.57 | 17.49 | 17.67 | 17.85 | 18.04 | 18.22 | 18.41 |
|---|---|---|---|---|---|---|---|---|
| $E_g$ | 0.22 | 0.17 | 0.12 | 0.11 | 0.11 | 0.10 | 0.09 | 0.0 |

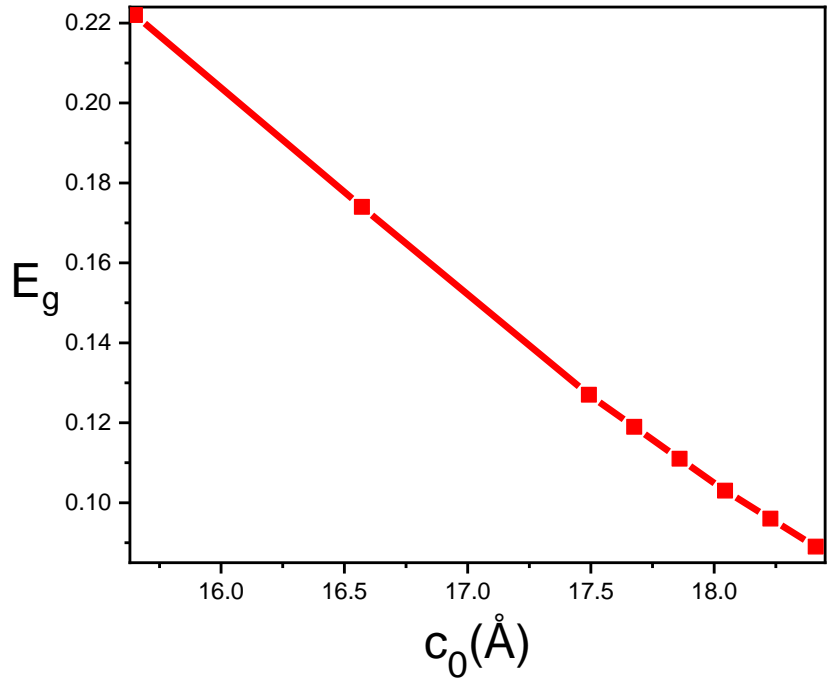

*Fig.4: the variation of the Gap energy as a function of the lattice parameter $c_0$.*

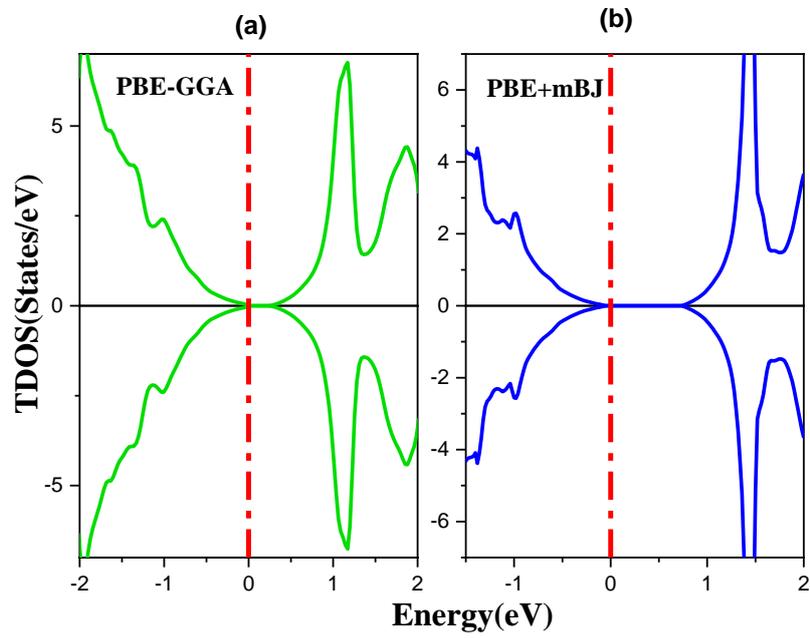

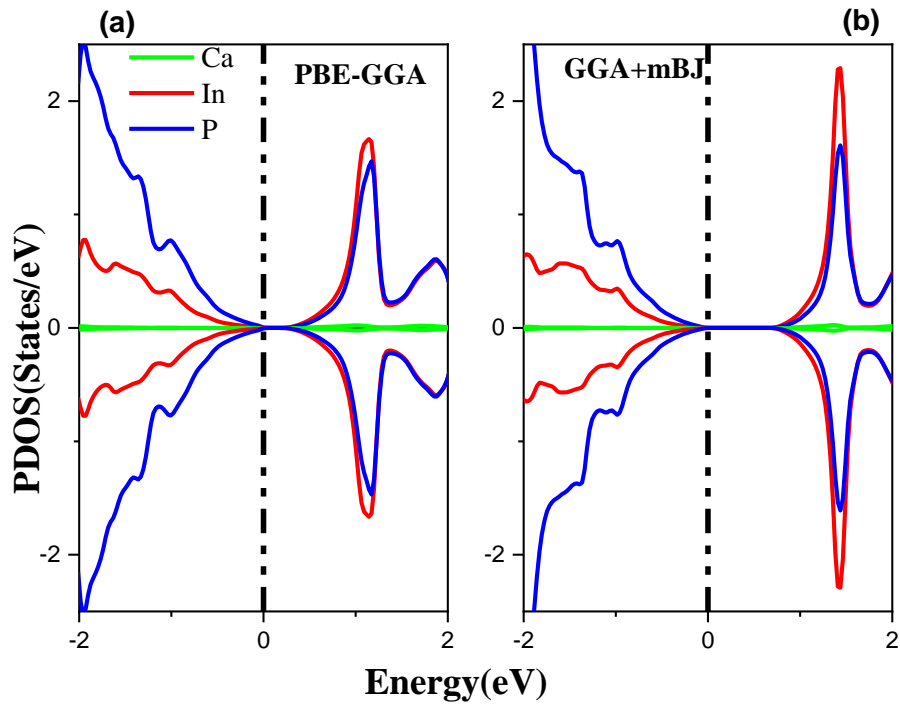

Fig. 5: The total (a) and partial (b) Density of States of Ca(InP)$_2$ using PBE-GGA and GGA+mBJ approaches.

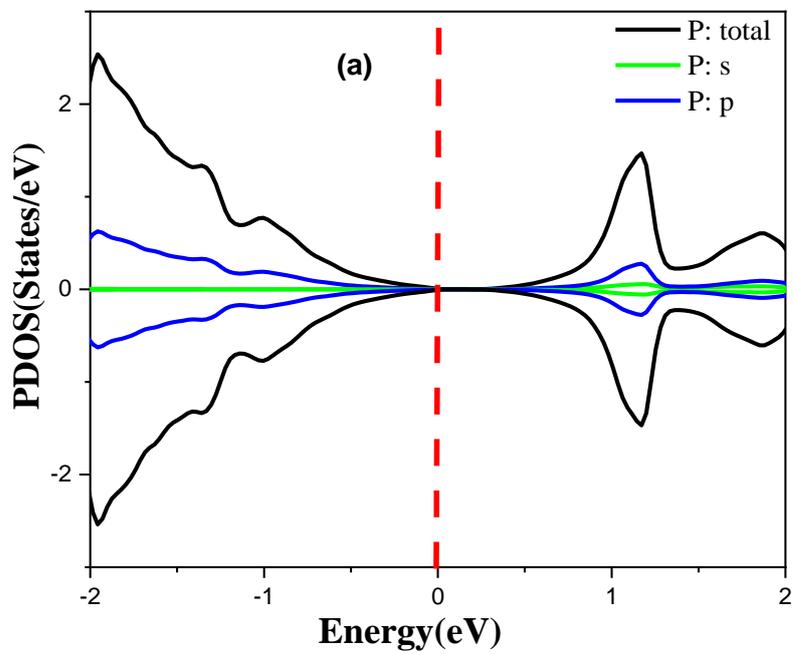

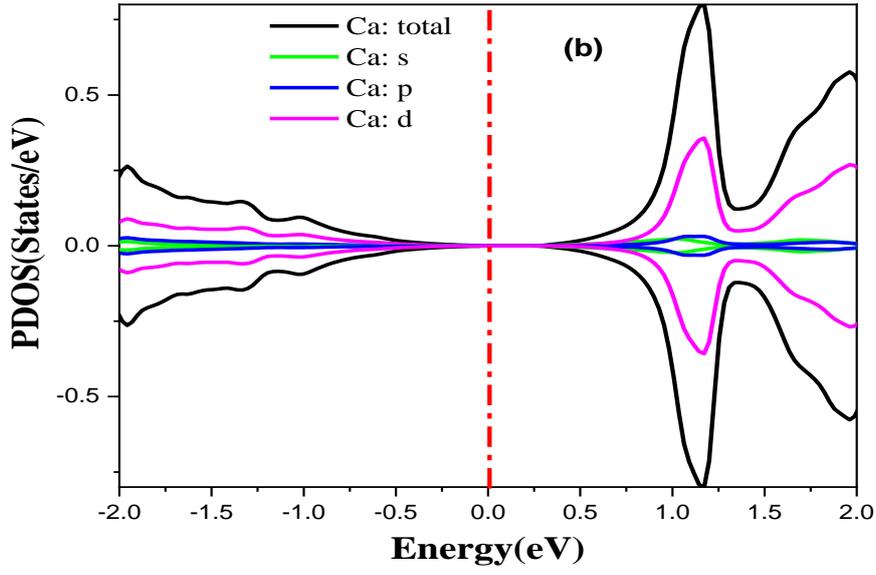

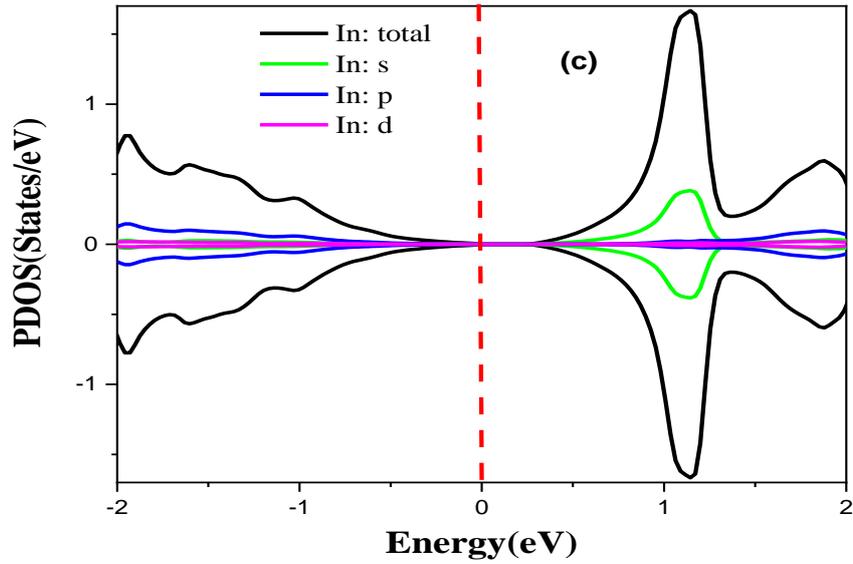

*Fig. 6: (a) The Total DOS of P and Decomposed DOS of P-s, P-p, (b) The Total DOS of Ca and Decomposed DOS of Ca-p, Ca-d, (c) The Total DOS of In and Decomposed DOS of In-s, In-p.*

### 3.3. Thermodynamic properties

To obtain the thermodynamic properties of Ca(InP)$_2$. According to the quasi-harmonic Debye model [31,32], wherein the Gibbs function

$$G(P,V,T) = E(V)PV + F_{Vib}(\Theta(V),T) \qquad (3)$$

Where E(V) is the total energy per unit cell for Ca(InP)$_2$ Θ(V) is the Debye temperature, and F$_{Vib}$ is the vibrational Helmholtz free energy. The thermal properties are determined in the temperature range 0 to 1800 K and the pressure effect is studied in the 0 to 12 GPa range [33,34].

$$F = -K_B T \ln Z = \frac{1}{2}\sum_{qj} \hbar\omega_{qj} + K_B T \sum_{qj} \ln\left[1 - \exp\left(-\frac{\hbar\omega_{qj}}{K_B T}\right)\right] \quad (4)$$

$$S = \frac{\partial F}{\partial T} = \frac{1}{2T}\sum_{qj} -\hbar\omega_{qj} \coth\left[\frac{\hbar\omega_{qj}}{2K_B T}\right] - K_B \sum_{qj} \ln[2\sinh(\hbar\omega_{qj}/2K_B T)] \quad (5)$$

$$C_V = \left(\frac{\partial E}{\partial T}\right)_V = \sum_{qj} K_B (\hbar\omega_{qj}/K_B T)^2 \frac{\exp\left(\frac{\hbar\omega_{qj}}{K_B T}\right)}{\left[\exp\left(\frac{\hbar\omega_{qj}}{K_B T}\right) - 1\right]^2} \quad (6)$$

with (F) Helmholtz free energy, entropy (S), constant volume heat capacity (C$_V$), partition function (Z), wave vector (q) and band index (j). These are fundamental parameters for understanding the complex thermodynamic properties of physical systems.

The relation between volume and pressure at different temperatures is shown in Fig. 7 for Ca(InP)$_2$ It can be seen the volume decreases with rising pressure while the temperature does not influence this variation. In Fig. 8, the heat capacity at constant volume C$_v$ is plotted for various pressures we can note that with increasing temperature, C$_v$ rapidly increases at low temperature up to the value of 600k, where it becomes almost constant, it tends towards the Dulong-Petit limit. Entropy plays a crucial role in thermodynamics, characterizing the degree of disorder within a system. It also serves as the basis for the definition of high-entropy alloys and compounds, representing the key concept for the development of new high materials. Thermodynamic entropy is linked to formation and stability, as well as to phase transition. It is therefore extremely important to study the variation of entropy with temperature as shown in Fig. 9, a rapid increase in entropy can be observed in the temperature range between 0 and 1800 K for different pressure values from 0 GPa to 8 GPa, indicating an increase in disorder. Thus, for a pressure of zero GPa Ca(InP)$_2$ is more stable in terms of entropy, which can be found in Fig. 10. For the thermal conductivity is shown in Fig. 11. Thermal conductivity [35] was calculated from Slack's equation given as

$$\kappa_L = \frac{A_\gamma M \delta^{\frac{1}{3}} \theta_D^3}{\gamma^2 n^{\frac{2}{3}} T} \tag{7}$$

Where $A_\gamma$ is a constant dependent on the Gruneisen parameter $\gamma = \frac{3(1+v)}{2(2-3v)}$, ($A_\gamma = \frac{(2.43 \times 10^{-8})}{1 - \frac{0.514}{\gamma} + \frac{0.228}{\gamma^2}}$). and M is the average atomic mass in amu, V is the volume per atom and $\theta_D$ Debye temperature and n is the number of atoms in the primitive unit cell. The curve shows the decrease in thermal conductivity with the increase in temperature from 0 to 1800 K for the range pressure 0-12 GPa, this means that these materials have a limited capacity to conduct heat due to an increase in phonon scattering.

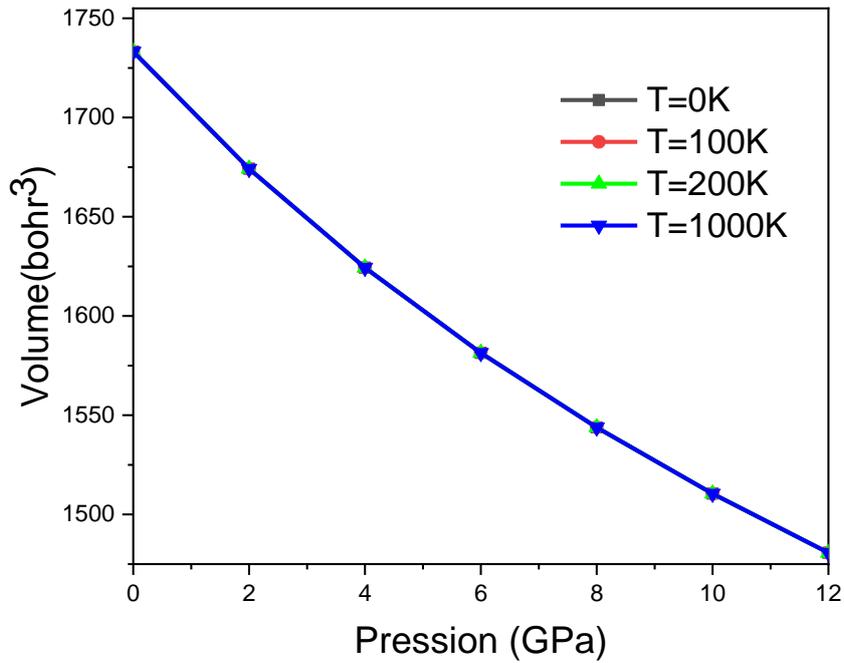

*Fig. 7: Volume variation with pression of $Ca(InP)_2$ at 0 K to 1000 K.*

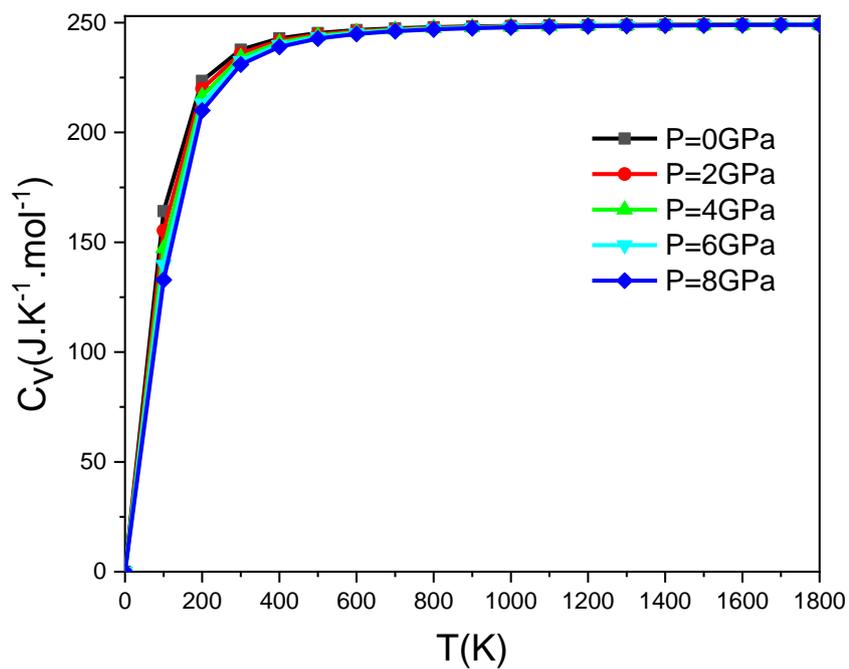

*Fig. 8: the heat capacity at constant volume as a function of temperature.*

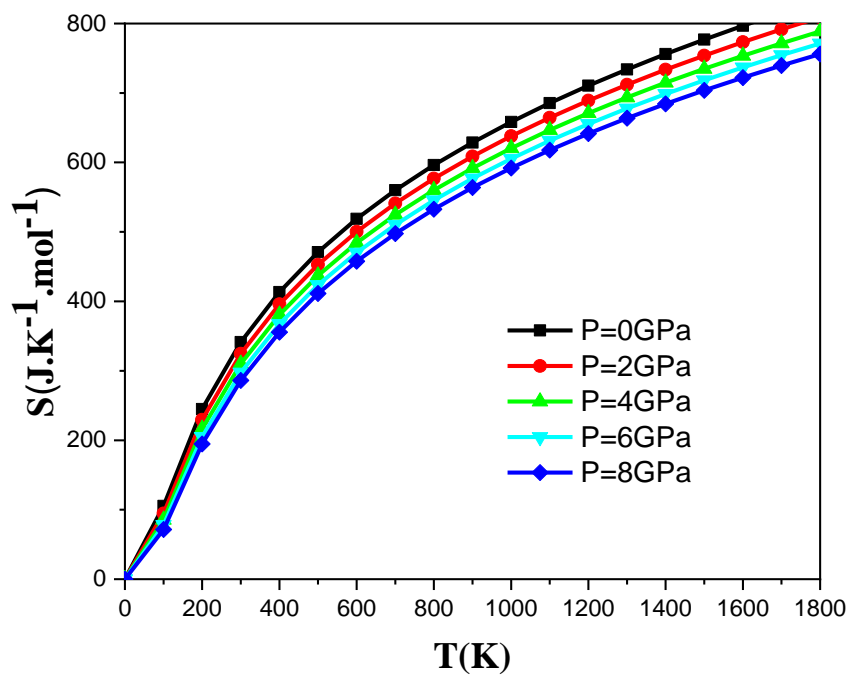

*Fig. 9: Calculated the entropy S (J.K$^{-1}$.mol$^{-1}$) as a function of temperature.*

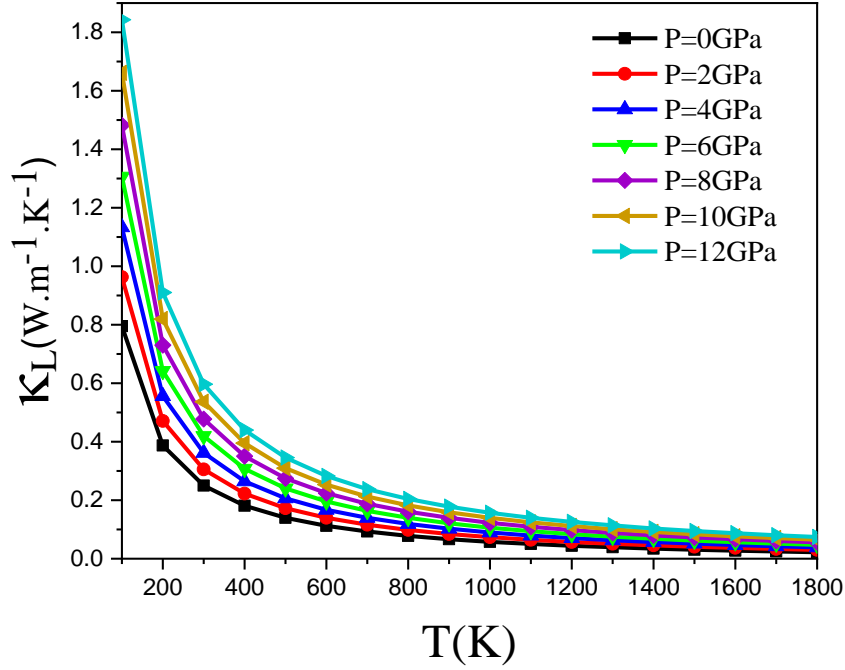

Fig. 10: Lattice thermal conductivity (κ_L) as a function of temperature.

### 3.3. thermoelectric properties

Thermoelectric properties refer to the characteristics of materials that determine their ability to transform thermal energy into electrical energy, and conversely, through the phenomenon known as the Seebeck effect. Key parameters include the Seebeck coefficient, electrical and thermal conductivity, and the thermoelectric figure of merit (ZT). For semiconductors and degenerate metals, the Seebeck coefficient is also calculated by [36]:

$$S = \frac{\pi^2}{3}\frac{k_B^2 T}{e}\left[\frac{1}{n}\frac{dn(E)}{d(n)} + \frac{1}{\mu}\frac{d\mu(E)}{d(n)}\right]_{E=E_F} \tag{8}$$

where n(E) = g(E) f(E) represents the charge carrier density as a function of electron energy, which is determined by multiplying the density of electronic states g(E) by the Fermi function f(E). Charge carrier mobility as a function of electron energy is represented by μ(E), and $k_B$ is the Boltzmann constant. Thus, S can be influenced by an increase in the density of states (dg(E)/dE) at the Fermi level. As shown in Fig. 11a, Seebeck coefficient values decrease with increasing temperature in the range between 100 K and 1800 K These results strongly support what we have obtained from the electronic study of Ca(InP)$_2$. This material is a P-type semiconductor. Thermal conductivity has two parts, the lattice contribution from phonon transport called the lattice thermal conductivity, $\kappa_L$, and the electrical contribution from electron/hole transport, $\kappa = \kappa_L + \kappa_e$, electronic thermal conductivity is based on the same transport function, which also drives the Seebeck coefficient and electrical resistivity. This applies particularly to a system with only one type of charge carrier, whether electrons or holes. in Figure 11.c, we observed a positive correlation between electrical conductivity and increasing temperature. This suggests that, as temperature increases, so does the material's ability to allow electron current to flow. This relationship can be explained by the increased thermal agitation of charge carriers, such as electrons, at higher temperatures. By increasing their kinetic energy, charge carriers have a greater probability of overcoming obstacles and collisions that may impede their movement, increasing electrical conductivity. This observation is of practical importance, as it suggests that the material may become more conductive at higher temperatures. An interesting trend in these analyses concerns the thermoelectric factor (ZT) as a function of temperature. Initially, the ZT value decreases with increasing temperature until it reaches 300 k. Then, a slight adjustment is made, and this difference in the curve corresponds exactly to the previously calculated Seebeck coefficient curve, as shown in Figure 11.a.

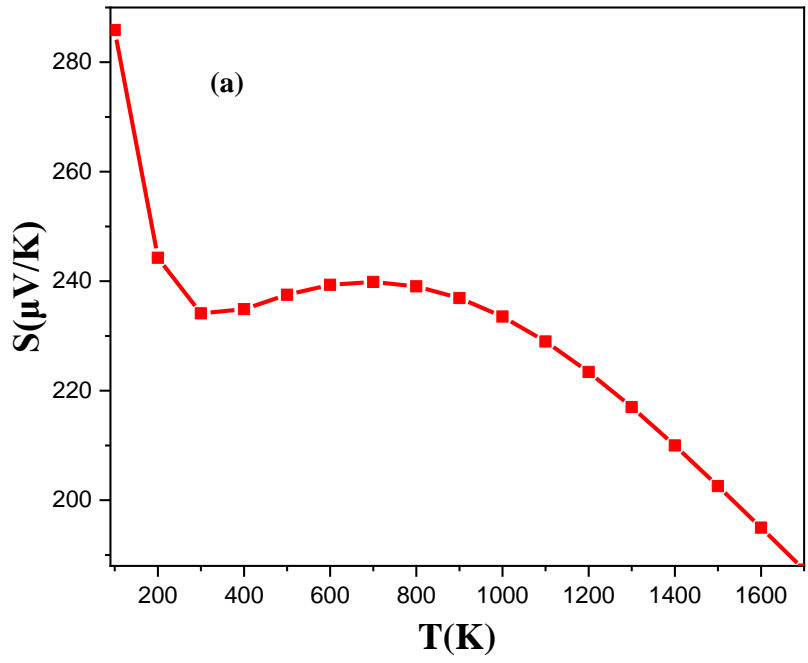
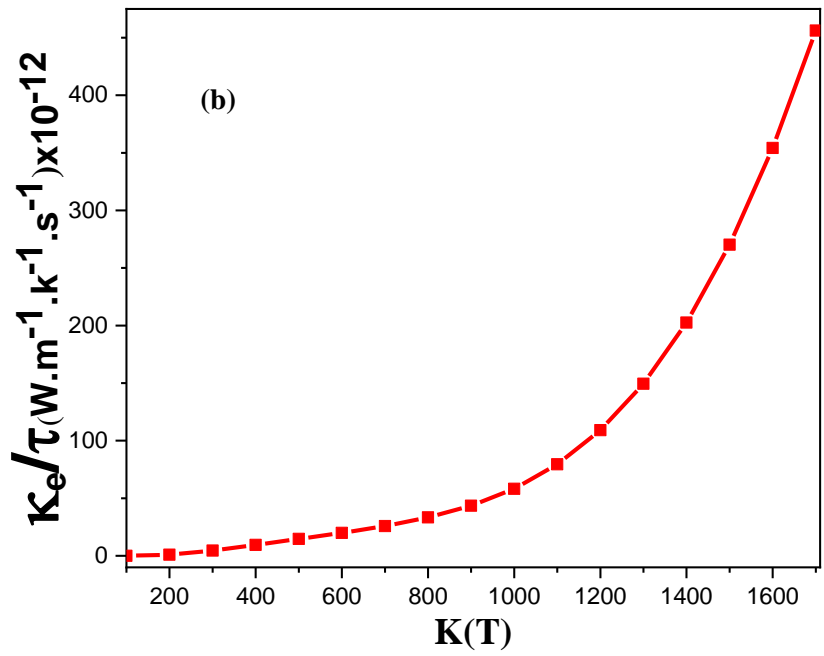

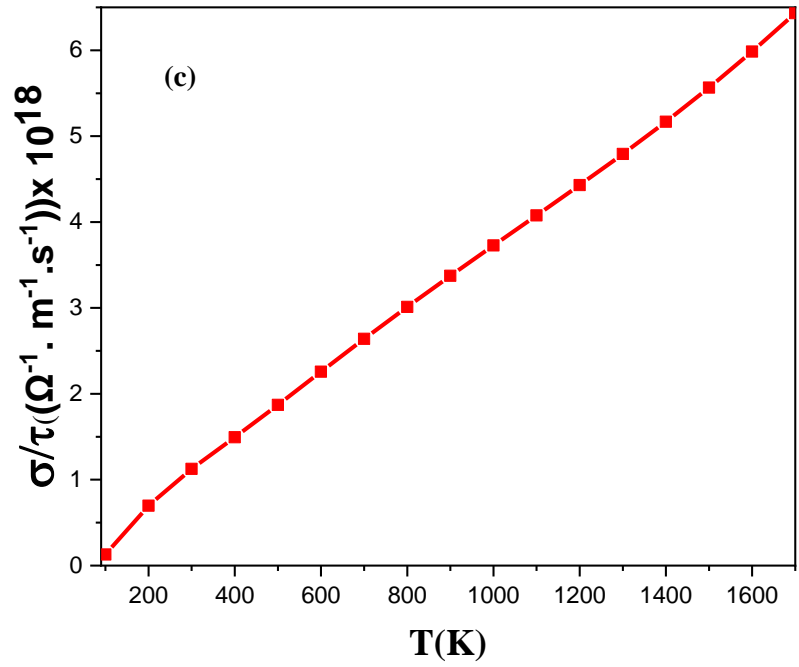

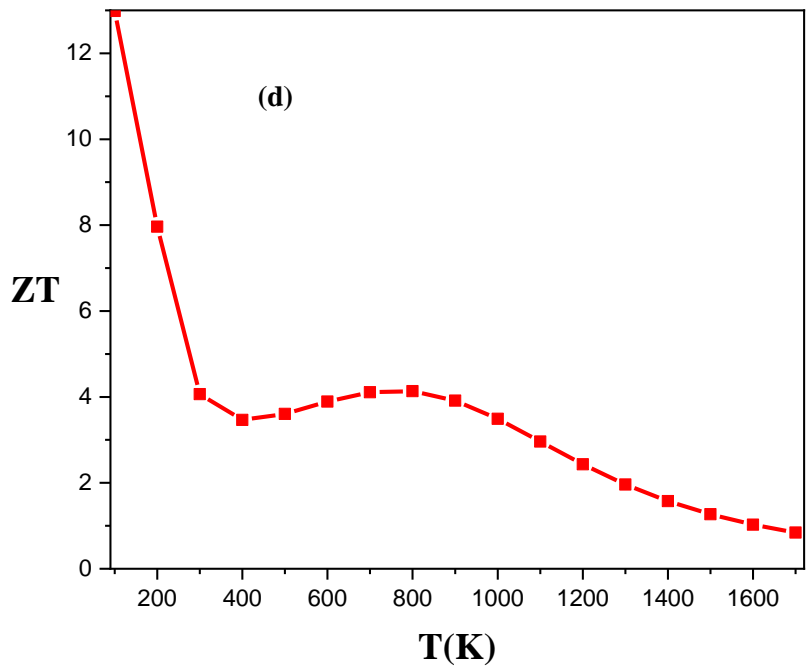

*Fig. 11: Calculated results of (a) the Seebeck coefficient, (b) electrical conductivity, (c) electrical conductivity and (d) figure of merit as a function of temperature.*

### 3.4. Optical properties

In this section, the optical response of the Ca(InP)$_2$ is made by the two approximations the PBE-GGA and GGA+mBJ. All optical properties are determined by the complex dielectric function given by:

$$\varepsilon(\omega) = \varepsilon_1(\omega) + \varepsilon_2(\omega) \qquad (9)$$

$$\varepsilon_1(\omega) = 1 + \frac{2}{\pi} P \int_0^\infty \frac{\omega' \varepsilon_2(\omega')}{\omega'^2 - \omega^2} d(\omega') \qquad (10)$$

And

$$\varepsilon_2(\omega) = \frac{8}{2\pi\omega^2} \sum_{nn'} \int |P_{nn'}|^2 \frac{dS_k}{\nabla_{\omega_{nn'}}(K)} \qquad (11)$$

with $\varepsilon_1(\omega)$ the real part measures the polarization of incident photons, and $\varepsilon_2(\omega)$ the imaginary part represents the absorption of incident light photons by the material. the studied materials are calculated in terms of optical parameters like conductivity, loss energy function, extinction coefficient, absorption coefficient and refractive index for Ca(InP)$_2$.

To investigate the impact of varying electronic band structure on optical properties, the imaginary part of the dielectric function, $\varepsilon_2(\omega)$, in the high-frequency range was evaluated using time-dependent first-order perturbation theory [37-39]. The imaginary part of the dielectric function $\varepsilon_2(\omega)$ represents transitions between occupied and unoccupied electronic states. Meanwhile, the real part of the dielectric function, $\varepsilon_1(\omega)$, has been calculated using the Kramers-Kronig relation [40]. The complex dielectric function is also used to determine other optical properties such as conductivity, refractive index n($\omega$), absorption coefficient $\sigma(\omega)$, energy loss function L($\omega$), reflectivity R($\omega$) and extinction coefficient k($\omega$). All these properties are closely related to each other through these equations [41], providing a better understanding of how variation in the electronic band structure can influence various aspects of the optical properties of the material under study.

$$\sigma(\omega) = \frac{\omega}{4\pi} \varepsilon_2(\omega) \tag{12}$$

$$n(\omega) = \sqrt{\frac{(\varepsilon_1(\omega)^2 + \varepsilon_2(\omega)^2)/2 + \varepsilon_1(\omega)^2}{2}} \tag{13}$$

$$\alpha(\omega) = \frac{2k\omega}{c} \tag{14}$$

$$L(\omega) = -Im(\epsilon(\omega)^{-1}) = \frac{\varepsilon_2(\omega)}{\varepsilon_1(\omega) + \varepsilon_2(\omega)^2} \tag{15}$$

$$R(\omega) = \frac{(n(\omega)-1)^2 + k(\omega)^2}{(n(\omega)+1)^2 + k(\omega)^2} \tag{16}$$

The refractive index in the energy range 0-12 eV for both approximations, the same behavior in both curves, as depicted in Fig 12.a, in the energy range 0-2 eV for PBE-GGA it has a maximum value of 2 eV along the x-axis and 4 eV along the z-axis and for GGA+mBJ it has the same maximum in energy value but the index refraction value is higher and after in the range between 4 eV to 13 eV the refractive index is decrease with increasing energy that means normal dispersion. An inverse relation between the imaginary dielectric function and energy loss, the function is also noticeable and visible in all the above-mentioned equations, the loss energy diagram in Figure 12.b, calculated by PBE-GGA, does not show a graph in the gap region as expected. However, naturally, after the gap, when the material reacts to the incident light, some of this light is lost, as shown in the figure, and its loss gap is larger for that of GGA+mBJ, which is in agreement with the electron gap, the maximum loss values are in the 12 to 13 eV range. For optical reflectivity of the single crystal of Ca(InP)$_2$ between 0 and 12 eV is shown in Fig.12.c. The real optical conductivity is displayed in part (d) of Figure 11, the highest values of real optical conductivity are 140 $\times 10^{15}$/s at 6 eV in PBE-GGA and 6.2 eV with GGA+mBJ. According to the PBE-GGA approximation, the imaginer optical conductivity is shown in Fig.12.e it has maximum values at 4.7 $\times 10^{15}$/s and 3.8 $\times 10^{15}$/s which are proportional to the energy of 2 eV and eV along the x and z axes, and for the GGA+mBJ approximation it has a maximum peak at 2.5 $\times 10^{15}$/s and 2 $\times 10^{15}$/s for the energy 6.5 eV and 5.5 eV respectively. We observed that the optical properties for the xx and xx axes are similar, which means that this isotropic material.

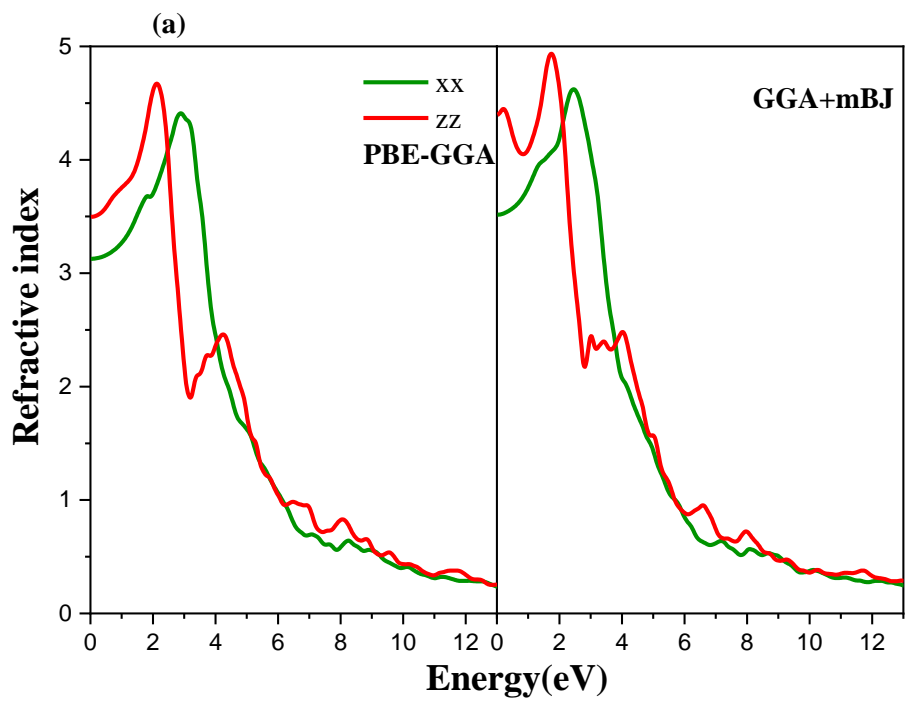

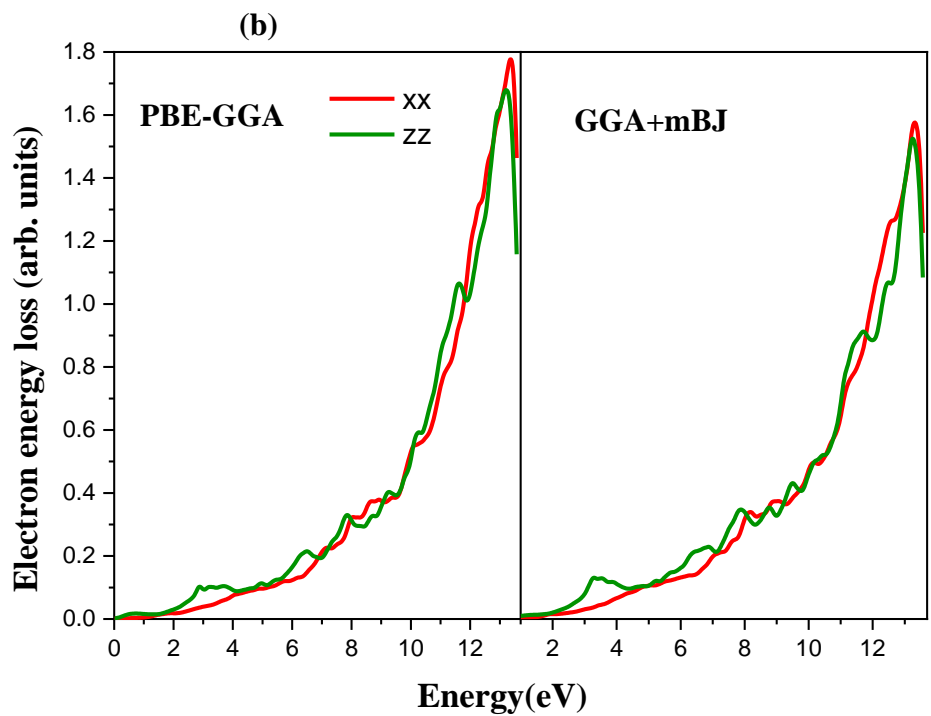

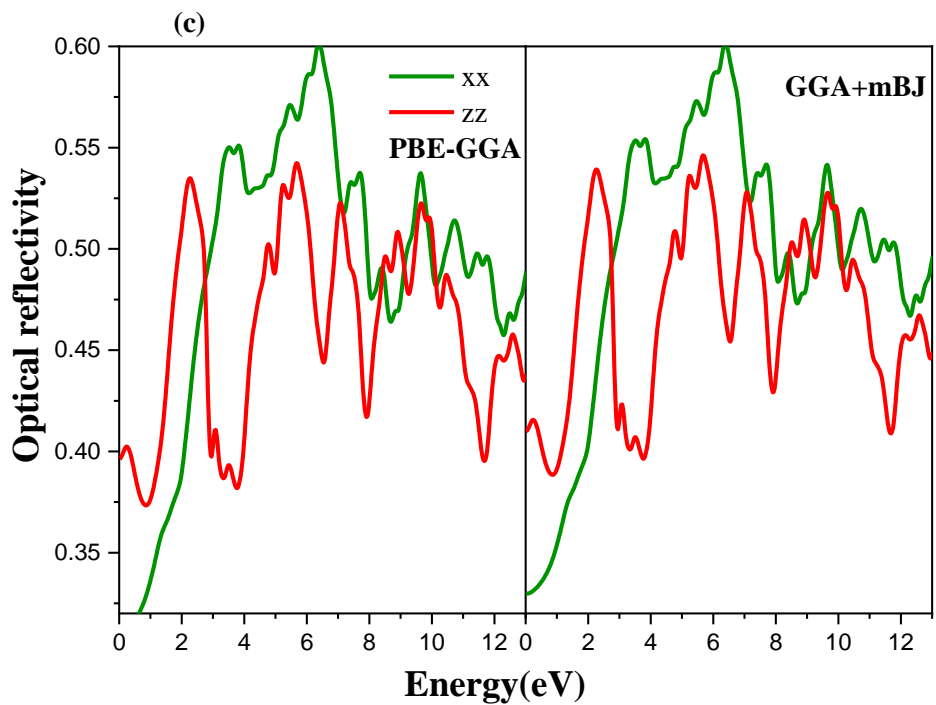

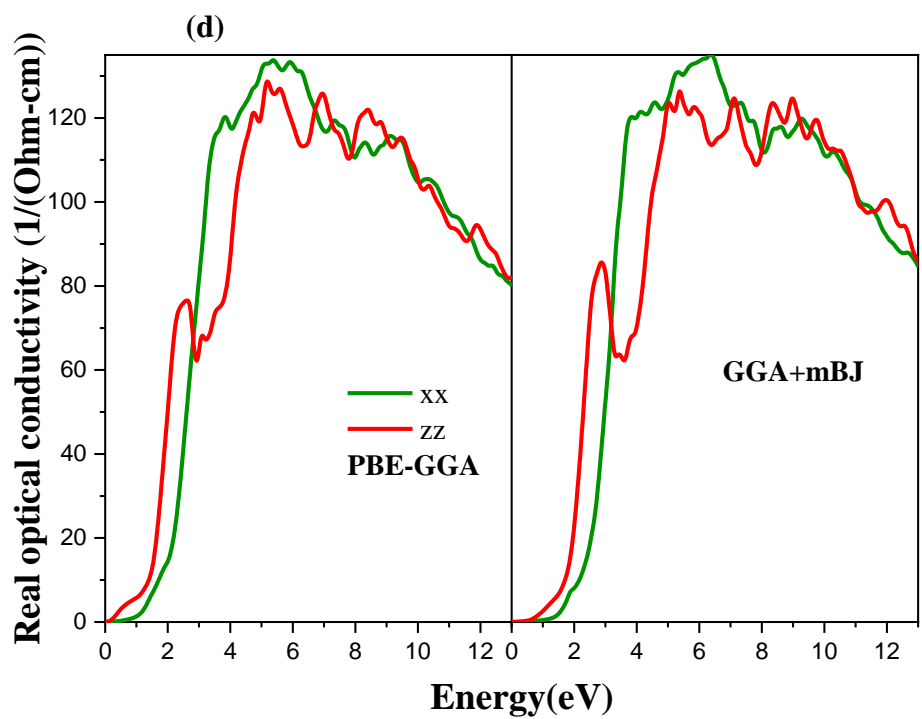

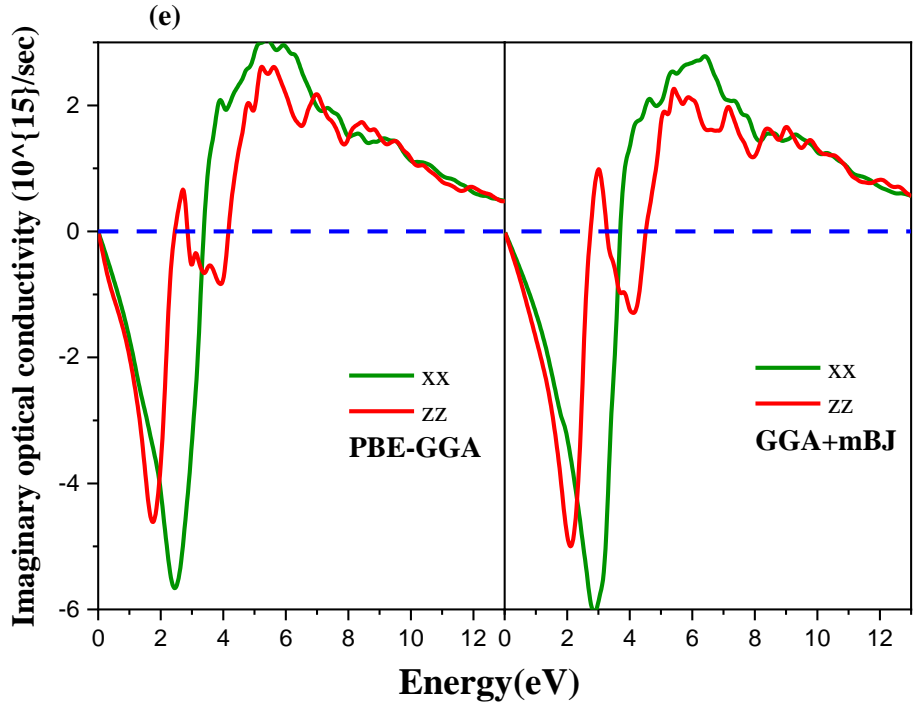

*Fig.12:* *Optical properties versus photon Energy (a) energy loss function, (b) reflectivity, (c) imaginary optical conductivity, (d) real optical conductivity and (e) refractive index.*

### 4. Conclusion

In conclusion, the electronic, thermodynamic and optical properties of Ca(InP)$_2$ were calculated based on DFT calculations with Wien2k using the PBE-GGA and GGA+mBJ approximations. Electronic results, including TDOS, PDOS and band structure diagrams, indicated conductive and semiconducting behavior through direct space at the Γ symmetry point for 0 eV and 0.645 eV using PBE-GGA and GGA+mBJ respectively. The total and partial densities of states are analyzed and studied in depth. An interesting trend in these analyses concerns the thermal factor, Seebeck coefficient and thermal conductivity. Optical parameters such as the energy loss function, reflectivity, imaginary conductivity, real conductivity and refractive index are producing promising results. With the support of experimental research, we hope this paper will inspire further studies.

However, despite advances in our understanding of the properties and application potential of Ca(InP)$_2$ and other two-dimensional materials, further research is essential to optimize synthesis and fabrication techniques. In addition, a better understanding of their fundamental properties is needed to fully exploit their potential in a wide range of practical applications, underlining the continuing importance of research in this area.

.